\documentclass{jpp}

\usepackage{graphicx}
\usepackage{natbib}
\usepackage{caption}
\usepackage{subcaption}

\ifCUPmtlplainloaded \else
  \checkfont{eurm10}
  \iffontfound
    \IfFileExists{upmath.sty}
      {\typeout{^^JFound AMS Euler Roman fonts on the system,
                   using the 'upmath' package.^^J}%
       \usepackage{upmath}}
      {\typeout{^^JFound AMS Euler Roman fonts on the system, but you
                   dont seem to have the}%
       \typeout{'upmath' package installed. JPP.cls can take advantage
                 of these fonts, if you use 'upmath' package.^^J}%
      }
  \else
  \fi
\fi

\ifCUPmtlplainloaded \else
  \checkfont{msam10}
  \iffontfound
    \IfFileExists{amssymb.sty}
      {\typeout{^^JFound AMS Symbol fonts on the system, using the
                'amssymb' package.^^J}%
       \usepackage{amssymb}%

      }{}
  \fi
\fi

\ifCUPmtlplainloaded \else
  \IfFileExists{amsbsy.sty}
    {\typeout{^^JFound the 'amsbsy' package on the system, using it.^^J}%
     \usepackage{amsbsy}}
    {\providecommand\boldsymbol[1]{\mbox{\boldmath $##1$}}}
\fi

\newsavebox{\astrutbox}
\sbox{\astrutbox}{\rule[-5pt]{0pt}{20pt}}

\title[Post-Disruptive Runaway Electron Beam]{Post-Disruptive Runaway Electron Beam in COMPASS Tokamak}

\author[M. Vlaini\'c et al.] 
{M\ls i\ls l\ls o\ls s\ns V\ls L\ls A\ls I\ls N\ls I\ls C$^{1,2}$%
  \thanks{Email address for correspondence: milos.vlainic@ugent.be},\ns
J.\ns M\ls L\ls Y\ls N\ls A\ls R$^2$,\ns
J.\ns C\ls A\ls V\ls A\ls L\ls I\ls E\ls R$^2$,\break
V.\ns W\ls E\ls I\ls N\ls Z\ls E\ls T\ls T\ls L$^2$,\ns
R.\ns P\ls A\ls P\ls R\ls O\ls K$^{2,3}$,\ns
M.\ns I\ls M\ls R\ls I\ls S\ls E\ls K$^{234}$,\break
O.\ns F\ls I\ls C\ls K\ls E\ls R$^{2,4}$,\ns
J.-M.\ns N\ls O\ls T\ls E\ls R\ls D\ls A\ls E\ls M\ls E$^{1,5}$\break
\and t\ls h\ls e\ns C\ls O\ls M\ls P\ls A\ls S\ls S\ns T\ls e\ls a\ls m}

\affiliation{$^1$Department of Applied Physics, Ghent University, Ghent B9000, Belgium\\[\affilskip]
$^2$Institute of Plasma Physics AS CR, Prague 18200, Czech Republic\\[\affilskip]
$^3$Faculty of Mathematics and Physics, Charles University, Prague 12116, Czech Republic\\[\affilskip]
$^4$Faculty of Nuclear Sciences and Physical Engineering, Czech Technical University, Prague 11519, Czech Republic\\[\affilskip]
$^5$Max Planck Institute for Plasma Physics, Garching D-85748, Germany}

\pubyear{2015}
\volume{???}
\pagerange{???--???}
\date{?; revised ?; accepted ?. - To be entered by editorial office}
\begin{document}

\maketitle

\begin{abstract}
For ITER-relevant runaway electron studies, such as suppression, mitigation, termination and/or control of runaway beam, obtaining the runaway electrons after the disruption is important. In this paper we report on the first achieved discharges with post-disruptive runaway electron beam, entitled ``runaway plateau'', in the COMPASS tokamak. The runaway plateau is produced by massive gas injection of argon. Almost all of the disruptions with runaway electron plateaus occurred during the plasma current ramp-up phase. Comparison between the Ar injection discharges with and without plateau has been done for various parameters. Parametrisation of the discharges shows that COMPASS disruptions fulfill the range of parameters important for the runaway plateau occurrence. These parameters include electron density, electric field, disruption speed, effective safety factor, maximum current quench electric field. In addition to these typical parameters, the plasma current value just before the massive gas injection surprisingly proved to be important.
\end{abstract}

\section{Introduction}

As the tokamak concept developed in the last 50 years and advanced towards the ITER design, numerous challenges occurred and many were solved. One of the remaining tasks is control or mitigation of Runaway Electrons (RE) in ITER after the disruption. Estimations from codes predict RE with several tens of MeV to carry up to 70\% of pre-disruptive plasma current \cite[p. S178]{Hender2007}. As deposition of runaway electron beam can be highly localised, it could severely damage plasma facing components and blanket modules of ITER.

The electron is said to run away, when the collisional drag force acting on it becomes smaller than the accelerating force coming from the toroidal electric field $E_{tor}$. There are three main mechanisms for the runaway generation: a) Dreicer (primary) mechanism \cite[]{Dreicer1959,Dreicer1960}; b) hot-tail mechanism \cite[]{Smith2008}; c) avalanche (secondary) mechanism \cite[]{Rosenbluth_Putvniski}. However, there is a theoretical limit for the electrical field, so called critical field $E_{crit}$, under which RE cannot be produced by these mechanisms \cite[]{Connor_Hastie}. The toroidal electric field $E_{tor}$ in ITER during the stable discharge will be under the $E_{crit}$ threshold, making the controlled ITER plasma void of the RE. On the other hand, if disruption occurs, the electron temperature $T_e$ would drop during Thermal Quench (TQ), and thus plasma electric resistivity $\eta$ would increase. $E_{tor}$, being proportional to $\eta \, j$, will rise dramatically during the Current Quench (CQ), because the current density $j$ drops much slower than the electric resistivity increase due to the vessel electromagnetic field penetration time. This increase of the field will first induce runaway seeds that will then be multiplied enormously by the avalanche effect. In the ITER disruption scenarios, the avalanche multiplication factor could be as large as $10^{22}$ \cite[table 5]{Hender2007}, forming an electron beam that could threaten ITER's first wall structure. Following the above outline, ITER should be equipped with a proper suppression and/or mitigation technique dedicated to the RE control. Thus, achieving post-disruptive RE beam is one of the first significant steps for COMPASS towards the ITER-relevant runaway suppression/mitigation studies.

The COMPASS tokamak \cite[]{COMPASS2006} is a small-size experimental fusion device with major radius $R_0=0.56$ cm and minor radius $a=0.23$\,cm. Toroidal magnetic field $B_{tor}$ is in $0.9 - 1.25$\,T range and plasma current $I_p$ can reach up to $330$\,kA. Electron densities are flexible and are typically of order of magnitude of $10^{19}-10^{20}$\,m$^{-3}$. Plasma shaping varies from circular and elliptical to single-null D-shaped ITER-like plasmas. The typical pulse length is $0.4$\,s, although the low current circular discharge with RE can last almost $1$\,s. Furthermore, flexibility of various plasma parameters (e.g. shaping, densities, plasma current, etc.) combined with significant, but still safe, runaway population make COMPASS suitable for runaway models validation and scaling towards ITER.

In contrary to large tokamaks (e.g. JET, ITER), where most of the RE are produced during the disruption \cite[]{Martin1996,Yoshino1999,Gill2000}, in small and medium size tokamaks RE are created either during the current ramp-up or the flat-top phase \cite[]{Esposito2003,PaprokNew} when $n_e$ is low and/or $E_{tor}$ is high enough. Additionally, the present COMPASS maximum value for $B_{tor}$ is $1.25$\,T, while various observations noted that getting the post-disruptive RE spontaneously is not possible if $B_{tor}$ is under $\approx 2$\,T \cite[]{Martin1996,Yoshino1999,Gill2002}. The $B_{tor}$-limit is the most probable reason for the lack of post-disruptive runaway observations in COMPASS. Therefore, size of the COMPASS and its maximum $B_{tor}$ could discard this facility from the ITER-relevant runaway suppression/mitigation research. Nevertheless, some of optimism can be found in experiments in which the post-disruptive RE were achieved with high-Z Massive Gas Injection (MGI) \cite[]{Yoshino1999, Gill2002, Hollmann2013} or high-Z pellet injection \cite[]{Yoshino1999,Hollmann2013}, since some of these experiments had $B_{tor}$ lower than $2$\,T. Moreover, a detailed study of $B_{tor}$-limit as a function of amount of Ar injected was performed recently in JET \cite[]{ReuxPSI2014}, where post-disruptive RE were observed even for $B_{tor}=1.2$\,T. Therefore, Ar injection was used to trigger the first post-disruptive RE in COMPASS.

The paper is organised as follows: in section \ref{In2Exp}, the experimental setup used for the experiments and demonstration of runaway plateau observation is presented. In section \ref{Res}, general runaway parameters are reported, followed by the injection and disruption details. The section is finalised with the discharge analyses of the parameters important for the plateau occurrence. In section \ref{Dis}, the results presented in section \ref{Res} are discussed. Finally, in section \ref{C_FW} conclusions and future perspectives are addressed.

\section{Introduction to Experiment}\label{In2Exp}

\subsection{Experimental Setup}

\begin{figure}
  \centering
    \includegraphics[width=0.5\textwidth]{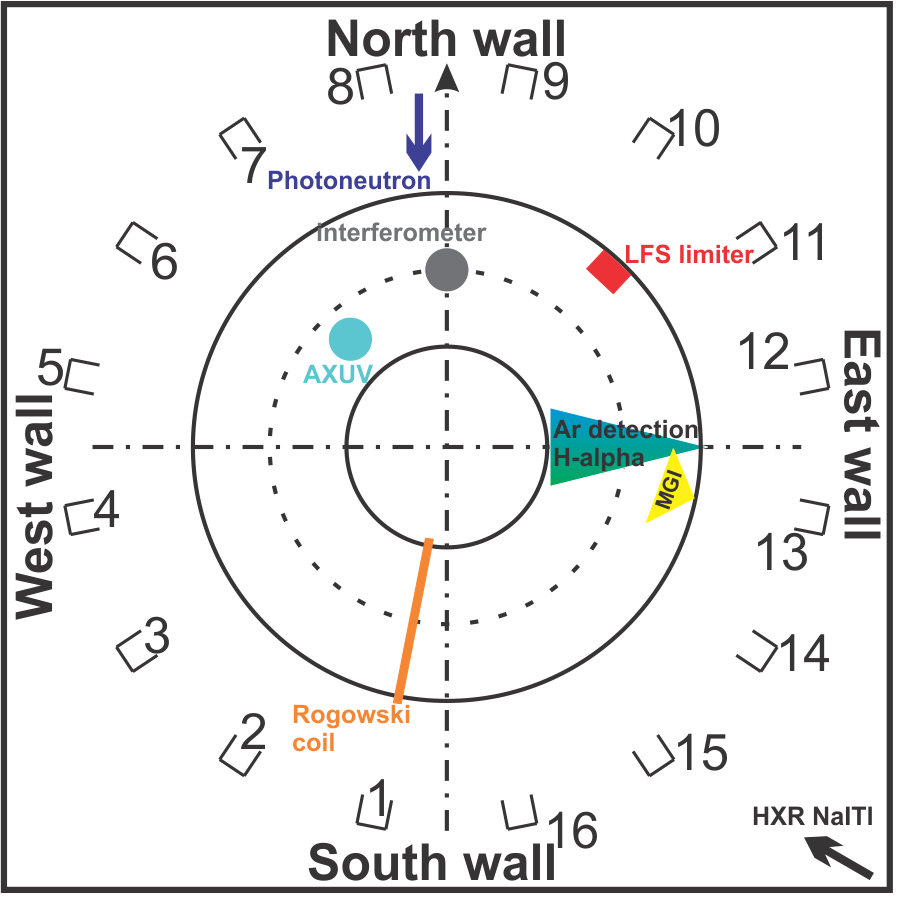}
  \caption{Principal diagnostics used for the RE plateau studies.}\label{ExpSetup}
\end{figure}

In all discharges described in this paper the plasmas were circular - limited by the carbon High Field Side (HFS) wall - with additional carbon Low Field Side (LFS) limiter for the inner wall vessel protection (see Fig.~\ref{ExpSetup}). Typical magnetic field $B_{tor}$ was $1.15$\,T and plasma currents $I_p$ at the moment of the gas injection varied from $40$ to $140$\,kA. The electron density $n_e$ was relatively low ($0.8-2.2 \times 10^{19}$\,m$^{-3}$) for maximising the runaway generation. Schematics of the experimental setup used in the experiments presented in this article is shown in Fig.~\ref{ExpSetup}. In this article, measurements of runaway losses will be presented from PhotoNeutron (PN) detector located nearby the north wall and NaI(Tl) scintillator for Hard X-Ray (HXR) detection located at the south-east part of the tokamak hall. Both detectors are approximately $5$\,m from the vessel. Photoneutrons with energy of few MeV are observed with the ZnS(Ag) neutron detector embedded in a plastic matrix. Beside the neutrons, the PN detector is suspected to be sensitive to the strong fluxes of HXR, although the detector is shielded by $10$\,cm of Pb. HXR are measured with unshielded NaI(Tl) scintillation detector, where the signal is amplified with a photomultiplier tube and the energy range is approximatelly from $100$\,keV to few MeV. Furthermore, the low energy photon radiation measurements will be presented from $H_{\alpha}$ detector and the bolometry. $H_{\alpha}$ detector is located radially at the east part of the tokamak vessel. AXUV photodiodes, located at the north-west part of the tokamak vessel, with photon energy response from $7$\,eV to $10$\,keV are used for bolometric measurements.

MGI of argon was achieved using a solenoid valve, located on the east side of the tokamak. The solenoid gas valve is connected to the vessel through two stainless steel tubes: the first one is $20$\,cm long and has an inner diameter of $4$\,mm, while the second one is $40$\,cm long and has an inner diameter of $6$\,mm. This non-negligible tube length implies a delay between the time of valve opening and the time at which the argon puff starts to interact with the plasma, i.e. roughly the time at which the gas enters the vacuum vessel. The delay is estimated to be approximately $1$\,ms taking into account a mean velocity of approximately $400$\,m$/$s for argon gas in vacuum at $300$\,K. 

The Ar flow rate $dN/dt$ through the injection system was evaluated experimentally as a function of the back pressure $p_{back}$ and with linear dependence as follows:

\begin{equation}
	\frac{\mathrm{d}N}{\mathrm{d}t} = (9.5511\,p_{back}-1.0083)\,10^{20},
\end{equation}

where $p_{back}$ is in bars and $dN/dt$ is in particles$/$s. The pressure $p_{back}$ used for the plateau discharges were $2.4$ and $1.2$\,bar, corresponding to particle flow rates of $2\times 10^{21}$ and $10^{21}$\,particles/s, respectively. The valve is roughly estimated to be open $2$\,ms, better knowlegde of gas valve performance will be available soon by the installation of a fast opening and more reliable valve. 

Since there is a pipe between the valve and the tokamak vessel, the duration time for Ar to enter the vessel is larger then the opening time of the valve and this Ar puff duration in the tokamak vessel will be estimated here. For the two aforementioned back pressures, $2.4$\,bar and $1.2$\,bar, the manufacturer gives a flux through the solenoid valve at standard conditions of $100$ and $50$\,Pa\,m${^3}/$s corresponding to flow rates of $2.4\times10^{22}$ and $1.2\times10^{22}$\,particles$/$s at $300$\,K, respectively. Therefore, assuming a constant flow rate of the solenoid gas valve with the increase of pressure in the stainless steel pipe and neglecting the flow rate through the injection system in the tokamak, one can calculate the number of particles that fill the pipe and that will be puffed in the tokamak later on. Remembering that the valve stays open for about $2$\,ms, one can find that there will be about $5\times10^{19}$ and $2.5\times10^{19}$ particles for $2.4$\,bar and $1.2$\,bar, respectively. Notice that these numbers are much smaller than the total number of particles that can be stored in the pipes at $2.4$\,bar and $1.2$\,bar ($8\times10^{22}$ and $4\times10^{22}$ respectively), justifying the assumption of constant flow rate through the solenoid valve. Now, knowing the flow rate through the injection system and the number of particles in the pipes, one can give an estimation of what the Ar puff duration in the tokamak vessel is: $25$\,ms for $2.4$\,bar and $12.5$\,ms for $1.2$\,bar. The runaway plateau created in this manner lasted from $2.5$ to $10$\,ms.

\subsection{Plateau Observation}

\begin{figure}
  \centering
  		\begin{subfigure}{0.48\textwidth}
            \caption{\#8585}\label{ExampleA}
		    \includegraphics[width=\textwidth]{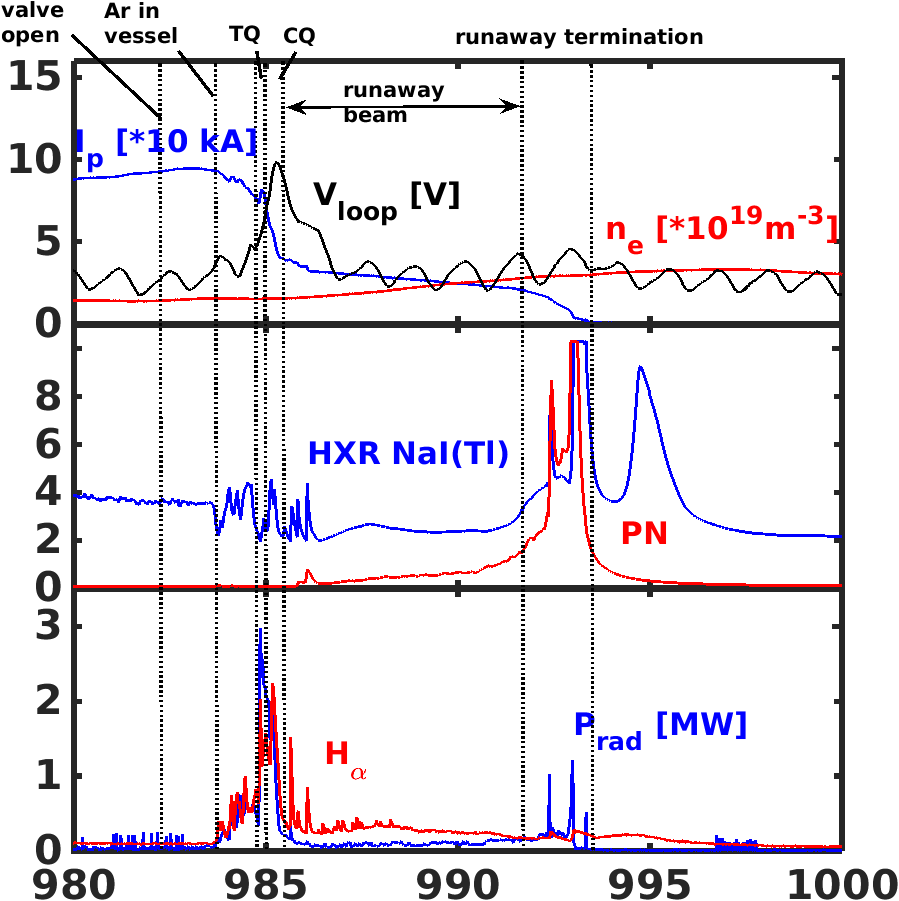}
    	\end{subfigure}
    	\begin{subfigure}{0.48\textwidth}
            \caption{\#8668}\label{ExampleB}
		    \includegraphics[width=\textwidth]{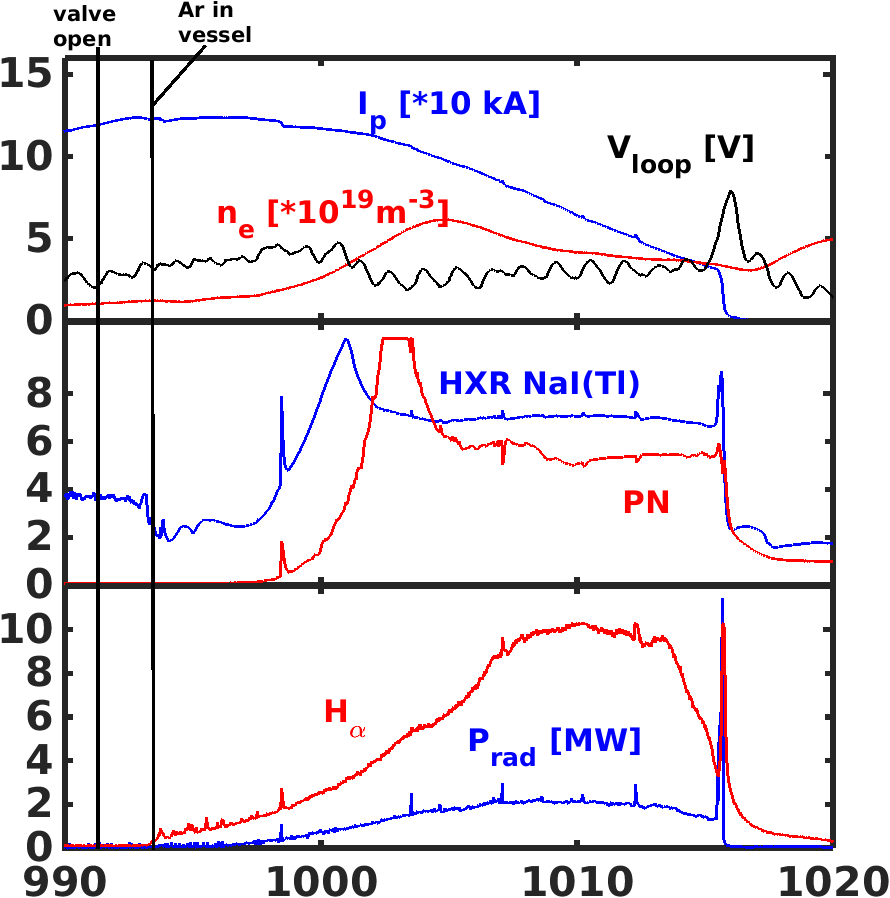}
    	\end{subfigure}
    \caption{Time evolution of the COMPASS discharge \#8585 as an example for runaway beam (a) and \#8668 as an example for slow $I_p$ decay (b) both initiated by the MGI. Plasma current $I_p$, electron density $n_e$ and loop voltage $V_{loop}$ are plotted on the top, in the middle HXR and PN signals are showing RE losses on the wall, while the $H_{\alpha}$ and $P_{rad}$ measurements are showing the radiation losses from plasma on the bottom (N.B. The y-axis are different for the bottom plot of the two discharges).}\label{Example}
\end{figure}

An example of a typical COMPASS discharge with MGI generated runaway plateau is shown in Fig.~\ref{ExampleA}, together with slow $I_p$ decay for comparison in Fig.~~\ref{ExampleB}. Fig.~\ref{ExampleA} shows the plateau discharge \#8585, when Ar puff starts to cool down the plasma, $I_p$ starts to drop and plasma radiation increases. After approximately $2$\,ms (this delay will be justified in the next section), TQ occurs and almost all plasma energy is radiated. At the same time the HXR measurement shows relatively low peaks in half saturated state and PN signal is rather low, meaning that high energy RE created during the discharge initial phase are still confined. Then, during the CQ, $E_{tor}$ is increased and boosts runaway production creating and amplifying the runaway beam. After the CQ the runaway beam carries non-zero current called $I_{RE}$, lasting for few milliseconds as one can see from the top graph in Fig.~\ref{ExampleA}. Finally, the RE beam terminates with the loss seen in HXR and PN signals, while there are almost no $H_{\alpha}$ radiation and radiated power $P_{rad}$ from plasma proving the runaway plateau existence. On the other hand, the COMPASS discharge \#8668 (see Fig.~\ref{ExampleB}) displays an example of the slow radiative decay with MGI on COMPASS resembling to $I_p$ ramp-down, for which no TQ and CQ (a typical sign of fast disruption) are observed. In this discharge plasma radiates on a long time scale ($\approx 20$\,ms). Although HXR and PN signals show presence of released RE, we shall not consider this as the runaway plateau, because the $I_p$ current is mainly driven by the thermalised plasma and not by the runaway beam, as one can see from strong $H_{\alpha}$ emission. Notice that the difference in $H_{\alpha}$ and $P_{rad}$ measurements makes the distinction between the runaway plateau and the slow radiative $I_p$ decay. The former one has a relatively low radiation level after the disruption, indicating that there is only cold plasma beside the runaway population.

\begin{figure}
  \centering
    \includegraphics[width=0.95\textwidth]{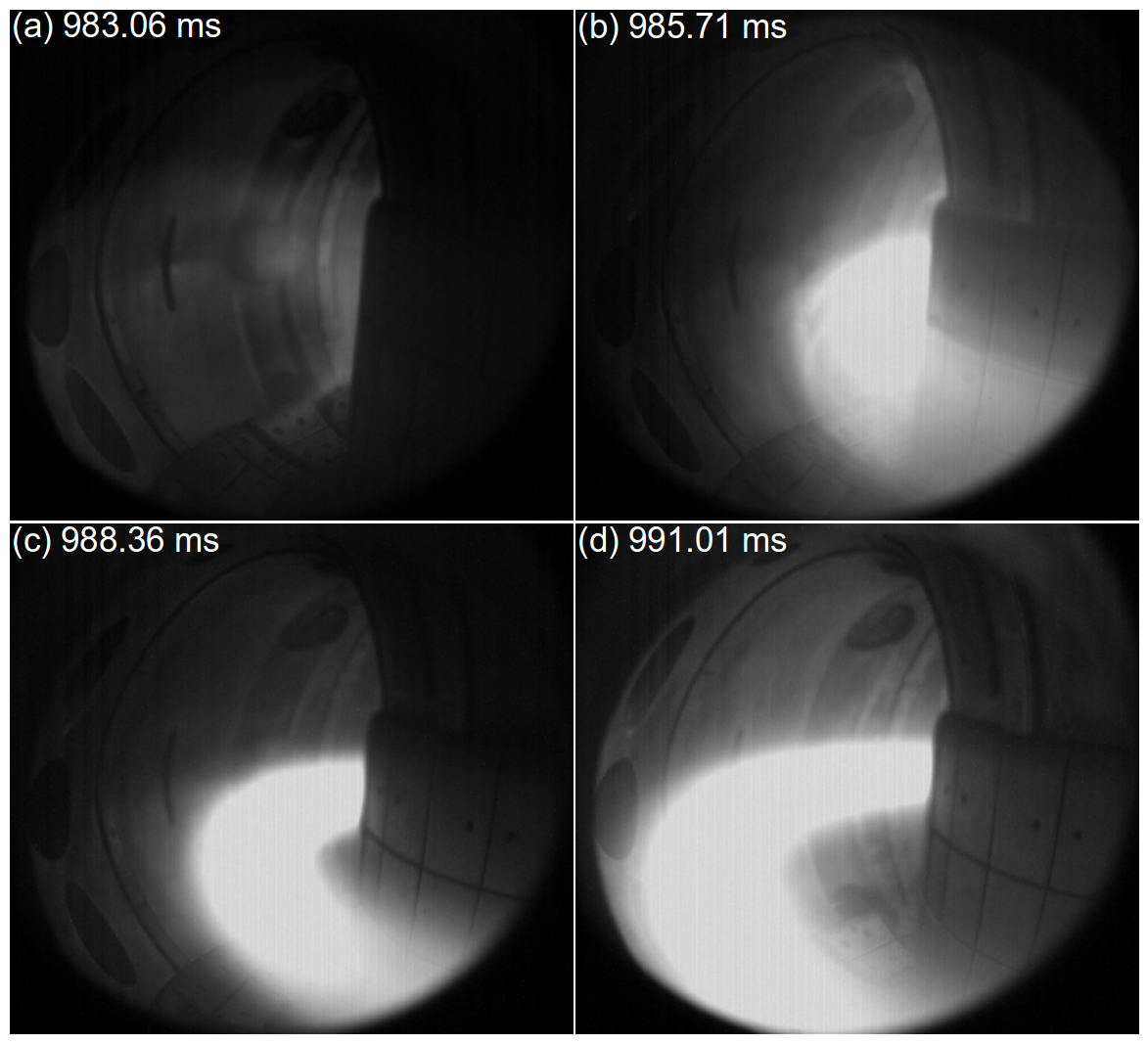}
  \caption{Visual observation of the RE beam with visible camera for discharge \#8585: (a) before Ar reaches vessel, (b) formation of the RE beam on HFS, (c) RE beam and (d) RE beam drifts towards LFS.}\label{Camera}
\end{figure}

Supplementary to the previous description of the RE plateau, the observation of RE beam with visible camera is displayed in Fig.\ref{Camera} for discharge \#8585. The creation and localisation of the beam are well visible.

\section{Results}\label{Res}

Out of 137 discharges performed during the COMPASS RE campaign, Ar puff was used in 39 discharges where only 5 discharges ended in spontaneous disruption. Out of the 39 discharges, 14 had the RE plateau after the Ar puff, while 9 resulted in slow radiative $I_p$ decay, similar to a ramp-down. The remaining 11 discharges ended in a typical COMPASS disruption, i.e. without any RE.

Based on these observations, all discharges with the Ar puff can be classified as:
\begin{enumerate}
	\item STRONG (RE plateau) - $I_{RE} > 5$\,kA
	\item WEAK (RE plateau) - $I_{RE} < 5$\,kA
	\item SLOW (radiative current decay) - plasma current slowly decreases in the similar manner as a ramp-down phase
	\item ZERO (RE plateau) - ``typical'' disruption for COMPASS with no RE remaining or generated after the disruption
\end{enumerate}

\begin{figure}
  \centering
  		\begin{subfigure}{0.48\textwidth}
  		 	\caption{All four classes.}\label{CompareA}
		    \includegraphics[width=\textwidth]{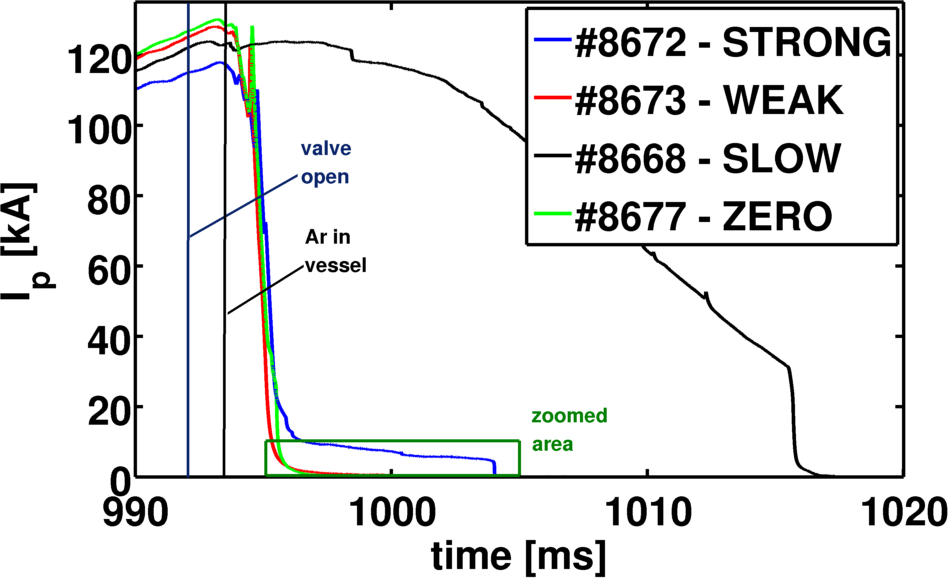}
        \end{subfigure}
    	\begin{subfigure}{0.48\textwidth}
            \caption{Zoomed area from (a).}\label{CompareB}
		    \includegraphics[width=\textwidth]{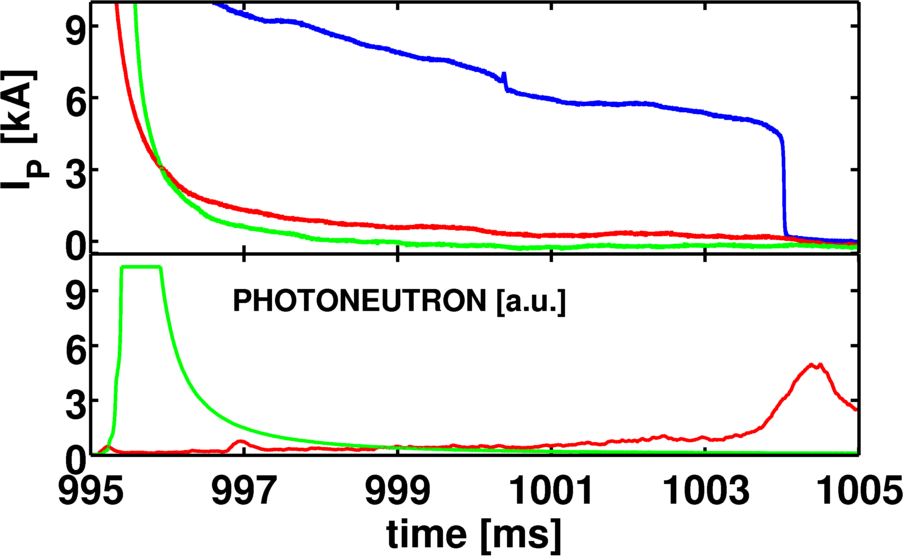}
    	\end{subfigure}
    \caption{Classification examples: (a) \#8672 for strong plateau, \#8673 for weak plateau, \#8668 for slow plasma current decay and \#8677 as an example of disruption without RE surviving or produced; (b) Zoomed region from part (a) for better observation of the difference between weak and zero plateau, as well as photoneutron signal for comparison of weak and zeros cases.}\label{Compare}
\end{figure}

An example of each class is shown in Fig.~\ref{Compare}, where Fig.~\ref{CompareB} is a zoom of Fig.~\ref{CompareA} to emphasize the difference between weak and zero plateau measurements. Although these two cases could seem identical at the first sight, the PN signal confirms release of the RE after the disruption in the weak case (b) and their loss during the disruption in the zero case (d). This classification is very important, as it will be used from now on throughout the paper. We shall now present the main results of the RE COMPASS campaign.

First, we have estimated the typical maximum runaway energy for the analyzed discharges to be $10-15$\,MeV by taking into account the electron acceleration due to electric field with the synchrotron radiation losses only, as suggested in \cite{Solis2010,Yu2013}.

Second, since the RE before or during the disruption are more likely to be produced in the hottest center of the plasma~\cite[]{Gill2002}, the remaining and newly produced post-disruptive RE may have more peaked radial current profile than the pre-disruption $I_p$ profile. The peaking represents localisation of the plasma current $I_p$ around the magnetic axis and can be expressed through the internal inductance $l_i$, which is calculated by the EFIT reconstruction \cite[]{WebMagnCOMPASS} at COMPASS. It was observed that the $l_i$ value increases by only 5-45\% in comparison to measurements on JET~\cite[]{Loarte2011}. In addition to the modest $l_i$ rise, the normalised plasma pressure $\beta_n$ rises above $1.5$ for the same discharges and thus confirms that the overestimated $\beta_n$ as seen by EFIT \cite[]{Vlainic2014} is caused by the presence of RE.

\begin{figure}
  \centering
    \includegraphics[width=0.95\textwidth]{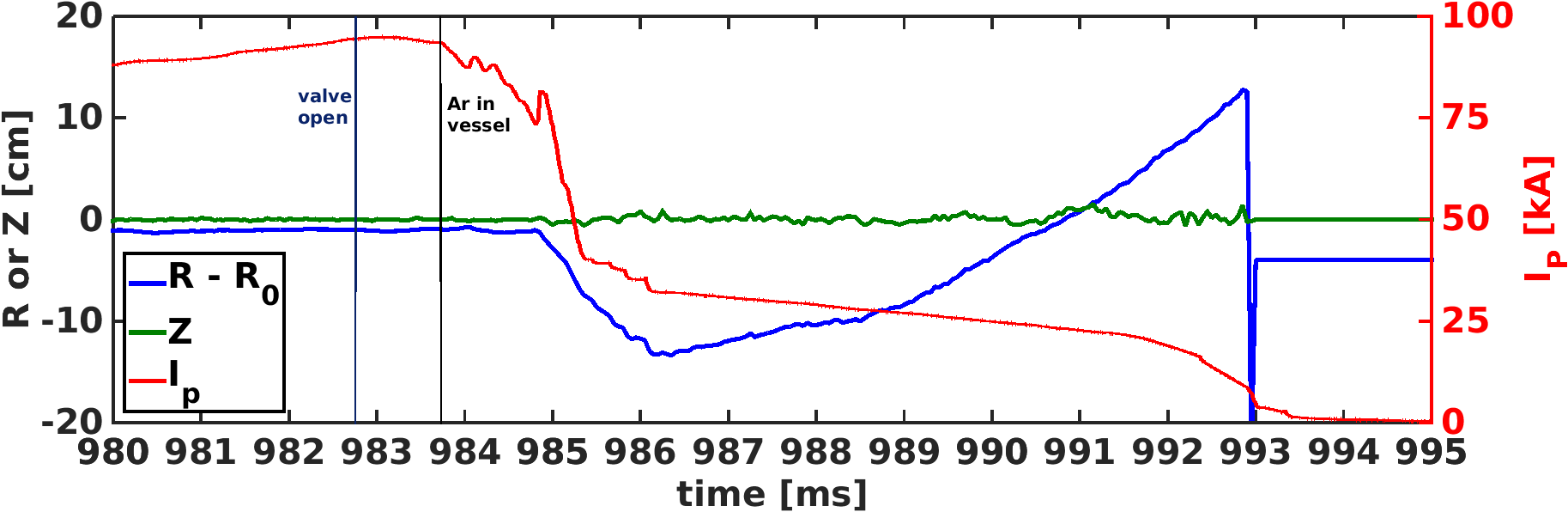}
  \caption{Time evolution of plasma vertical $Z$ and horizontal $R-R_0$ positions for the COMPASS discharge \#8585, associated with the plasma current $I_p$ barycentre. Positive values of $R-R_0$ mean that the plasma $I_p$ barycenter is closer to the LFS and positive values of Z mean that the plasma $I_p$ barycentre is closer to the top of the vessel.}\label{Movement}
\end{figure}

The inward motion (towards the HFS wall, negative $R-R_0$ values in Fig.~\ref{Movement}) of the post-disruptive plasma, followed by its return towards the vessel center is in agreement with the TFTR~\cite[]{Fredrickson2015} and Tore Supra~\cite[]{SLeps2009} observations. However, TFTR and Tore Supra feedback systems were able to stabilise the runaway beam, while presently in COMPASS the beam continues to shift outwards until its termination, as shown in Fig.~\ref{Movement}. Note that the outward shift is also visible in Fig.~\ref{Camera}. The vertical plasma position for the majority of cases is rather stable (an example being given in Fig.~\ref{Movement}), only in a few discharges some downward shifts were noticed.

\subsection{Disruption Generated by Argon}

As already mentioned, a solenoid valve was used to inject Ar gas into the plasma. Even though two different pressures were used ($2.4$ and $1.2$\,bar), no particular differences in runaway beam parameters were identified. The reason could be that the pressure was only varied by a factor of 2.

In devices larger than COMPASS, high-Z gas injection is used to trigger fast CQ in order to improve runaway generation \cite[]{ReuxPSI2014}. The plasma current quench rate $I_{\gamma}$:
\begin{equation}\label{Igama}
	I_{\gamma} = \frac{1}{I_p}\frac{\mathrm{d}I_p}{\mathrm{d}t}
\end{equation}
is the quantifying parameter for the CQ speed. The calculation of $I_{\gamma}$ values for disruptions with and without Ar puff was performed. No particular differences were observed between the discharges, as the majority of the $I_{\gamma}$ values are between $500-2500$\,s$^{-1}$ in both cases. The values implicate that the whole pre-disruptive $I_p$ is lost in $0.4-2$\,ms, which is the order of magnitude of the electromagnetic field penetration time of the COMPASS vacuum vessel ($\sim0.5$\,ms).

\subsection{Parametrisation of Runaway Plateau}

\begin{figure}
  \centering
    \includegraphics[width=0.5\textwidth]{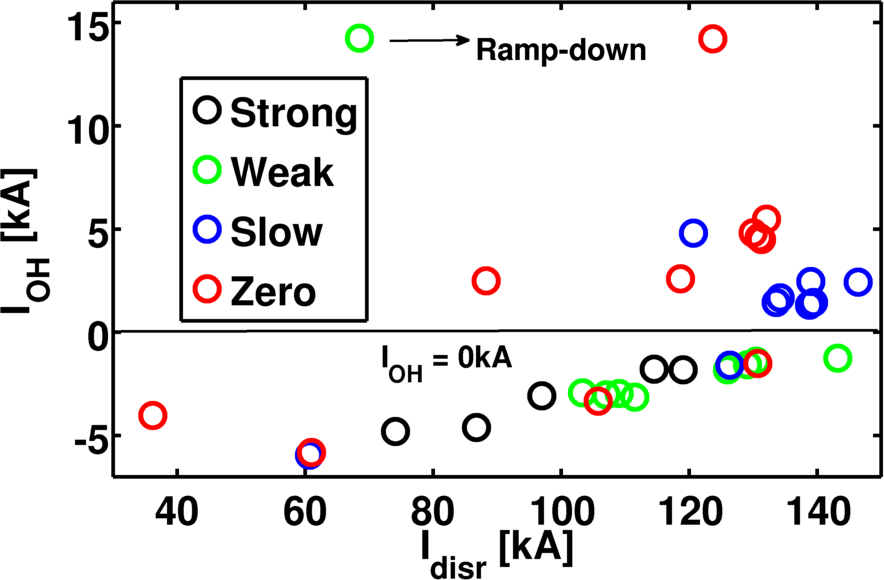}
  \caption{$I_{OH}$ as a function of the plasma current before the gas puff $I_{disr}$. Negative values of $I_{OH}$ correspond to the current ramp-up phase, while positive values represent the current flat-top and ramp-down phases.}\label{RampUp}
\end{figure}

The ohmic heating (OH) central solenoid current $I_{OH}$ - called MFPS in \cite[]{Havlicek2008} - will be used to indicate on the appearence time of the runaway plateau. For the RE discharges analyzed in the article $I_{OH}$ is negative during the current ramp-up phase followed by $I_{OH}$ at zero value for few milliseconds during the transition towards the current flat-top phase. For the rest of the discharge, i.e. current flat-top and ramp-down, it becomes positive and controlled by the feedback system \cite[]{Janky2014}. In Fig.~\ref{RampUp}, $I_{OH}$ $2$\,ms before TQ is plotted versus $I_p$ also taken $2$\,ms before the TQ and denoted as $I_{disr}$. The reason why exactly $2$\,ms are taken will be seen later in this section, but it can be explained as the time before Ar starts to cool down the plasma, displayed in Fig.~\ref{CompareA}. Also, later in the article, the measured parameters denoted with the index $disr$ (e.g. $E_{disr}$ and $n_{disr}$) are taken at the same time.

Fig.~\ref{RampUp} shows that only one weak plateau out of 14 plateau discharges did not appear during the ramp-up phase, but during the ramp-down phase. Hence, RE plateaus are more likely to be produced during the current ramp-up phase than during the flat-top. The ramp-down case requires further investigation in future experiments, as only one such discharge was observed.

\begin{figure}
  \centering
    \includegraphics[width=0.5\textwidth]{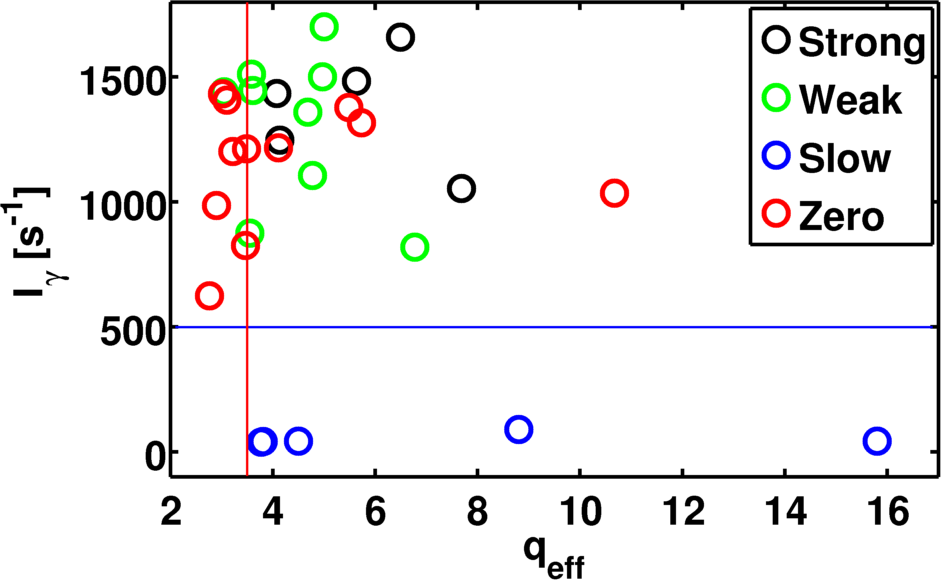}
  \caption{Plasma current quench rate $I_{\gamma}$ as function of $q_{ef\!f}$. Vertical red line corresponds to $q_{ef\!f}=3.5$ and horizontal blue line corresponds to $I_{\gamma}=500$\,s$^{-1}$.}\label{YoshinoSpace}
\end{figure}

\cite{Yoshino1999} did the first detailed parametrisation of disruptions with runaway occurence in JT-60U tokamak. According to his article, study of $I_{\gamma}$ versus $q_{ef\!f}$ is important for the plateau occurrence, where the effective edge safety factor $q_{ef\!f}$ for circular plasma is defined as:
\begin{equation}
	q_{ef\!f} = \frac{5a^2B_{tor}}{R I_p} \left[ 1+ \left(\frac{a}{R}\right)^2 \left( 1+\frac{(\beta_p+l_i/2)^2}{2} \right) \right].
\end{equation}
The internal inductance $l_i$ and the poloidal beta $\beta_p$ are taken from the EFIT reconstruction at the closest moment from the disruption. $I_{\gamma}$ is already defined in Eq.~\ref{Igama}. Fig. \ref{YoshinoSpace} shows $I_{\gamma}$ versus $q_{ef\!f}$ for the case of COMPASS. For all plateaus except the slow ones, $I_{\gamma}$ is between $500$ and $1800$\,s$^{-1}$ and $q_{ef\!f}$ is between $2.5$ and $8$. It is interesting to observe how majority of the zero disruptions are under $q_{ef\!f}=3.5$. Obviously, slow disruptions have significantly slower current decay than the rest of the discharges, their $I_{\gamma}$ values are under $100$\,s$^{-1}$.

\begin{figure}
  \centering
  		\begin{subfigure}{0.45\textwidth}
            \caption{}\label{NeBorderA}
		    \includegraphics[width=\textwidth]{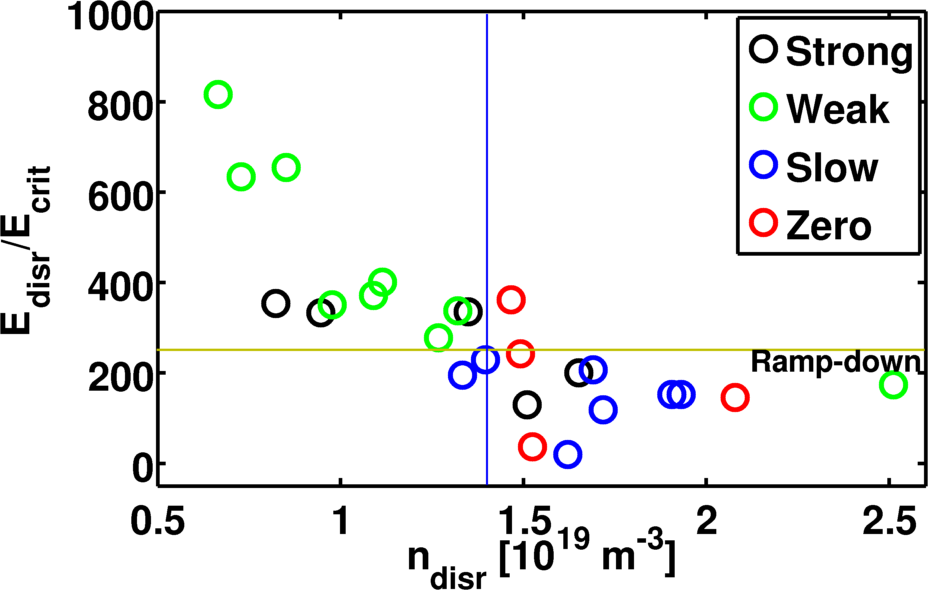}
    	\end{subfigure}
    	\begin{subfigure}{0.45\textwidth}
            \caption{}\label{NeBorderB}
		    \includegraphics[width=\textwidth]{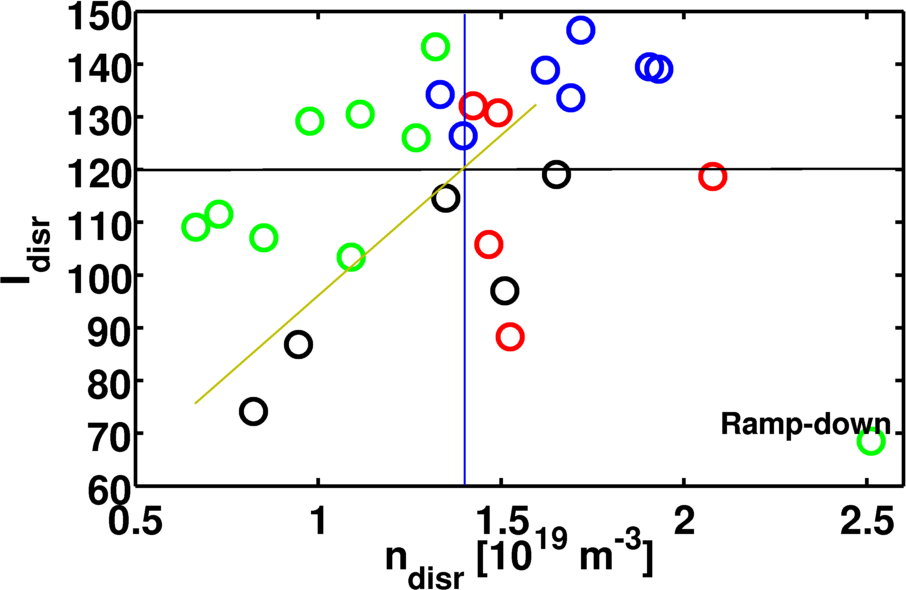}
    	\end{subfigure}
    \caption{Normalised electric field $E_{disr}/E_{crit}$ (a) and plasma current just before the MGI puff $I_{disr}$ (b) as function of the electron density $n_{disr}$. The vertical blue line matches $n_{disr} = 1.4 \times 10^{19}$\,m$^{-3}$. The horizontal red line in (a) corresponds to $E_{disr}/E_{crit} = 250$, while black line in (b) is for $I_{disr}=120$\,kA. The oblique green line in (b) represents the limit between the strong and the weak case.}\label{NeBorder}
\end{figure}

According to the theory the production of the RE is more intense for lower densities. Thus, $E_{disr}$ normalised to $E_{crit}$ and $I_{disr}$ are plotted as a function of the line averaged density $n_{disr}$ measured by the interferometer in Fig. \ref{NeBorder}. Approximately, the critical value of electron density for obtaining the runaway plateau seems to be $1.4 \times 10^{19}$\,m$^{-3}$. The ratio $E_{disr}/E_{crit}$ represents the relative strength of $V_{loop}$. Critical value of the  $E_{disr}/E_{crit}$ ratio on Fig.~\ref{NeBorderA} is around $250$ for the analyzed discharges. Figure~\ref{NeBorderB} shows that the strong plateaus are created for lower $I_{disr}$ value than weak plateaus, taking the same $n_{disr}$ value. In addition, no strong plateau is observed above $I_{disr} = 120$\,kA, while half of the weak ones have $I_{disr}$ above $120$\,kA.

\begin{figure}
  \centering
  		\begin{subfigure}{0.45\textwidth}
            \caption{}\label{IreA}
		    \includegraphics[width=\textwidth]{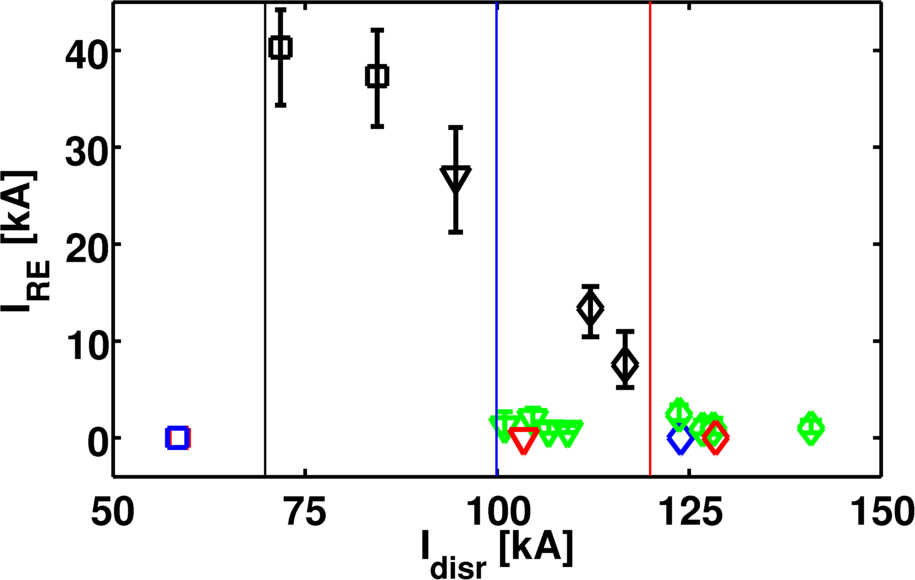}
    	\end{subfigure}
    	\begin{subfigure}{0.45\textwidth}
            \caption{}\label{IreB}
		    \includegraphics[width=\textwidth]{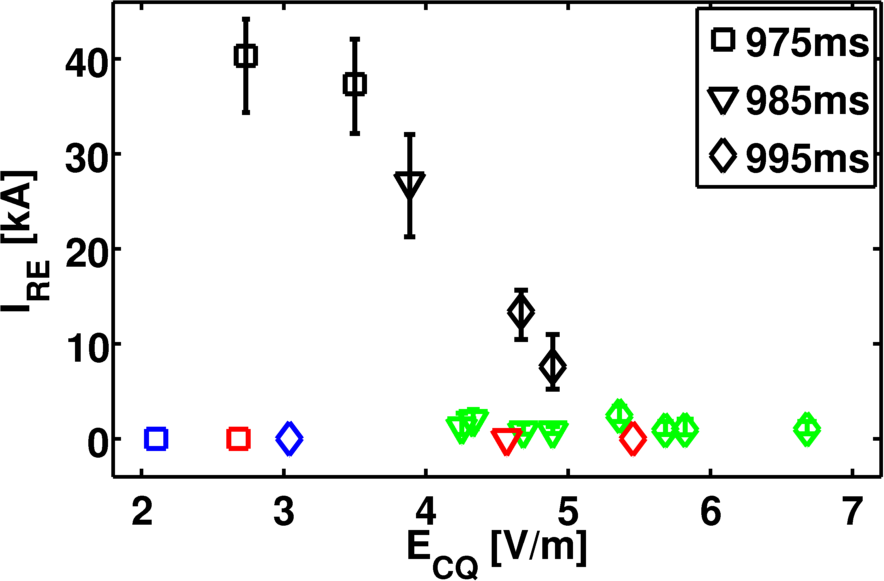}
    	\end{subfigure}
    \caption{Runaway beam current $I_{RE}$ as function of the pre-disruptive current $I_{disr}$ (a) and the maximum loop voltage during the current quench $E_{CQ}$ (b). The colors are kept the same as in previous figures, while different marker type corresponds to different times of the Ar injection: $975$\,ms (squares), $985$\,ms (triangles), $995$\,ms (diamonds). Symbols stand for a mean value of the $I_{RE}$, and error bars stand for maximum and minimum values of the $I_{RE}$. The vertical lines in (a) match $70$\,kA (black), $100$\,kA (blue) and $120$\,kA (red).}\label{Ire}
\end{figure}

Another parameter of interest is the current carried by the RE beam $I_{RE}$. The dependence of $I_{RE}$ on the $I_{disr}$ is shown in Fig.~\ref{IreA}, where only the ramp-up Ar MGI discharges are presented. The discharges in Fig.~\ref{Ire} are grouped by the time of Ar puff. Typical weak plateau $I_{RE}$ is between $0.5-3.5$\,kA, while $I_{RE}$ for strong plateaus decreases with $I_{disr}$ and time of the puff. Furthermore, for the Ar injections performed at $985$\,ms and $995$\,ms the upper limit of the $I_{disr}$ is indicated, namely $100$\,kA and $120$\,kA respectively. For values lower than these $I_{disr}$ values strong plateau seems to be produced, while above either weak plateau or no plateau occurred. In contrast to this, for the Ar injection at $975$\,ms the lower limit of $I_{disr}$ is observed for about $70$\,kA, under which no plateau was detected. Anyway, more statistics are required. The dependence of $I_{RE}$ on the maximum electric field $E_{CQ}$ during the current quench (Fig.~\ref{IreB}) has similar behavior like in Fig.~\ref{IreA}, as one could expect from the self-inductance effect between $I_{disr}$ and $E_{CQ}$. 

\section{Discussion}\label{Dis}

Even though the number of discharges devoted to runaway plateau studies was limited on COMPASS in the dedicated RE campaign, it was still possible to do comparative analyses. The results presented in the previous section are discussed in following order:
\begin{itemize}
	\item general characteristics on RE are outlined
	\item observed differences between discharges with and without the runaway plateau are reported
	\item issues on obtaining the RE plateau with Ar puff are discussed
	\item achieving strong plateau is commented
\end{itemize}

In~\cite{Loarte2011}, an increase of $l_i$ by factor of 2 to 3 has been reported for the RE plateau, which is significantly larger then the modest rise observed in COMPASS (0.05-0.45). This modest $l_i$ rise could mean that RE seeds generated during the ramp-up phase are occuring in larger relative area than for the case of JET. Similarly to Tore Supra~\cite[]{SLeps2009}, JET~\cite[]{Loarte2011} and TFTR~\cite[]{Fredrickson2015}, the inward motion of the RE beam is observed for the beginning of the plateau phase at COMPASS. 

Almost all (13 out of 14) generated RE plateaus were achieved for Ar puff in the ramp-up current phase, as for Tore Supra~\cite[]{SLeps2011}. Regarding disruptions itself, the CQ speed is one order of magnitude larger than for the case of JT-60U \cite[]{Yoshino1999}, where $I_{\gamma} > 100-200$\,s$^{-1}$ was reported as the plateau formation condition. Concluding that COMPASS has fast enough disruption for the plateau formation (see Fig.~\ref{YoshinoSpace}), but other factors - e.g. $B_{tor}$, $V_{loop}$, avalanching \cite[]{Rosenbluth_Putvniski} - are not fulfilled, explaining why the Ar MGI is necessary for COMPASS to obtain the runaway plateau.

Next, $q_{ef\!f}$ and its relation with $I_{\gamma}$ is one more important plasma characteristics for the plateau creation. In the case of JT-60U \cite[]{Yoshino1999}, on top of the $I_{\gamma}$ condition, $q_{ef\!f}$ has to be over $2.5$. From Fig.~\ref{YoshinoSpace} it is obvious that disruptions analyzed here are deep in the reported parameter region. However, there is an indication how the plateau condition for $I_{\gamma}$ and $q_{ef\!f}$ in COMPASS could be different from those observed in JT-60U. Anyhow, this is still to be investigated by enhancing the statistics.

As observed from Fig.~\ref{NeBorderA}, the limiting $n_{disr}$ for plateau to appear is around $1.4\times10^{19}$\,m$^{-3}$, which corresponds to $E_{crit}=0.0122$ V$/$m. The Dreicer mechanism is the most probable source of the post-disruptive production of RE at COMPASS, because the avalanching is expected to be important for the tokamaks with $I_P \gtrsim 1$\,MA~\cite[]{Rosenbluth_Putvniski}. However, from Fig.~\ref{NeBorderA} it is apparent how the strong plateaus are obtained for low $E_{disr}/E_{crit}$ values compare to the weak plateaus. This observation, at the first sight contra-intuitive, could be explained with the appearance of the avalanching effect, as avalanching is dominant runaway generation mechanism for lower $E_{disr}/E_{crit}$ assuming that the electron temperature profile remains unchanged~\cite[see][Fig.~10]{Nilsson2015}. Anyhow, this possibilty is still to be investigated. The Dreicer field $E_{D}$ is currently difficult to determine as the Ar injections were often too early, so that no Thomson scattering data were collected yet. For the cases plotted in Fig.~\ref{NeBorderB}, it seems that lower densities are necessary in order to achieve plateau for $I_{disr}$ above $120$\,kA, making $I_{disr}$ important parameter for the plateau production.

In the COMPASS case, it seems that the inverse dependence is recognised for the strong plateaus (Fig.~\ref{IreA}). The dependence of $I_{RE}$ on $I_{disr}$ from Fig.~\ref{IreB} looks almost the same as the one from Fig.~\ref{IreA}, as one would expect. This observation comes from the fact that the amplitude of induced $E_{tor}$ during the CQ is directly proportional to the $I_p$ before the disruption. Lower and upper boundary signs of plasma current for strong plateaus from Fig.~\ref{IreB} are not unique, these boundaries have been observed in JET by \cite{Gill2002}. In this article, the lower limit is assigned to low $E_{tor}$, while the upper limit is possibly connected to magnetic fluctuations. For COMPASS more discharges would be required to improve the statistics and find the two limits.

\section{Conclusion and Future Work}\label{C_FW}

Before this dedicated campaign, runaway plateau was never observed in COMPASS. As a matter of fact, there was scepticism concerning the possible plateau occurrence for any plasma condition, due to the COMPASS tokamak size, low $B_{tor}$ and the low plasma currents leading to relatively low electric field $E_{tor}$ during the disruption. Nevertheless, in this paper a clear demonstration of obtaining runaway plateau by MGI is reported. The RE plateau currents varied between $0.5$ to $40$\,kA, with duration from $2.5$ to $10$\,ms.

Argon injection disrupted discharges in COMPASS have been investigated in order to clarify the necessary conditions for runaway plateau production. It was found that the easiest way to produce the RE plateau was to inject Ar during the ramp-up of the plasma current. Furthermore, the typical COMPASS disruptions without RE can satisfy various parameters important for the runaway plateau creation (e.g. $n_e$, $V_{loop}$, $I_{\gamma}$, $q_{ef\!f}$) without Ar injection, thus high-Z MGI is probably required only for activating thermal quench to enhance runaway population. Unusually, for the discharges considered in the paper, it seems that the plateau generation also depends on plasma current during the Ar puff injection. Even though, the CQ after MGI induced disruption has a very short time, it is possible that avalanche mechanism is present in COMPASS during the runaway plateau formation.

More experiments need to be done in order to draw final conclusions on the definite conditions for the runaway plateau generation in COMPASS tokamak. From present knowledge we can conclude that some observations correspond to reports from larger tokamaks, although the amplitude is sometimes different. Indeed, this difference in the magnitudes could be important for scaling towards ITER.

The experiment presented here confirm that COMPASS is a tokamak suitable for various ITER-relevant runaway studies, such as:
\begin{enumerate}
	\item studies of runaway plateau termination - energy balances and timescales \cite[]{Loarte2011, Solis2014}
	\item improvements of the runaway beam mitigation
	\item testing the runaway control system
	\item benchmarking of the runaway models 
\end{enumerate}
Nonetheless, the scenario for inducing the runaway plateau is necessary before further ITER-relevant studies are performed. Presently, LUKE~\cite[]{LUKE} code is being used in collaboration with CEA for a better understanding of the physics behind the measurements.\\

The ``Joint Doctoral Programme in Nuclear Fusion Science and Engineering'' is acknowledged by the first author for supporting the studies. Next to thank is the project MSMT LM2011021 from which the COMPASS operation is supported. Then, the authors would like to acknowledge work of the WP14-MST2-9 research project team. The first author would like to thank to Francois Saint-Laurent for advising and sharing his experience with us, Jozef Varju for installing the injection system and Josef Havl\'i\v{c}ek and Michael Komm for fruitful discussions. Faculty of Nuclear Sciences and Physical Engineering (Czech Technical University) is also appreciated for lending us the HXR detectors.

This work has been carried out within the framework of the EUROfusion Consortium and has received funding from the Euratom research and training programme 2014-2018 under grant agreement No 633053. The views and opinions expressed herein do not necessarily reflect those of the European Commission.


\bibliographystyle{jpp}

\bibliography{biblio}

\end{document}